\documentclass{emulateapj}
\usepackage{natbib}
\usepackage{amsfonts}
\usepackage{amssymb}

\shorttitle{Accelerated Recombination}
\shortauthors{Jaiseung Kim and Pavel Naselsky}
\begin{document}
\title{Accelerated Recombination, and the ACBAR and WMAP data}
\author{Jaiseung Kim and Pavel Naselsky}
\affil{Niels Bohr Institute, Blegdamsvej 17, DK-2100 Copenhagen, Denmark}
\email{jkim@nbi.dk}
\submitted{Submitted to the Astrophysical Journal Letter} 

\begin{abstract}
We have investigated deviation from the standard recombination process, using the ACBAR 2008 and the WMAP 3 year data. In this investigation, we have considered the possibility of accelerated recombination as well as delayed recombination.
We find that accelerated recombination is as likely as delayed recombination, and there is some degeneracy between $\epsilon_{\alpha}$ and \{$n_s$, $\log[10^{10}A_s]$, $H_0$\}.
\end{abstract}

 
\section{Introduction}
The ionization history of the cosmic plasma is one of the most important part of the
modern cosmology. Based on well-known atomic physics principles, the theory of recombination of the cosmological hydrogen provides remarkable information about the most fundamental properties of the matter in the Universe through the anisotropy and polarization of the Cosmic Microwave Background (CMB). In this Letter, we discuss different modification of this  model and some of the  observational consequences of the nonstandard models of recombination, proposed in \cite{Delayed_recombination,Baryonic_recombination,Ionization_history,Baryonic_recombination2} (see also \cite{Constraint_Recombination} and the references therein). The early source of Ly-$\alpha$ photons may lead to the deviations from the standard recombination process, hence affecting the CMB anisotropy \citep{Delayed_recombination,Ionization_history}. 

The major distortions of the theoretically predicted power spectrum, which are related to acceleration or delay of the hydrogen recombination at the redshift $z_{rec}\simeq 1000$, are expected to be found in the CMB polarization \citep{Ionization_history}.  The temperature anisotropy of the CMB is less sensitive to the distortions of the recombination at $z=z_{rec}$. This is why the future PLANCK polarization data will provide a unique opportunity to test the nonstandard models of the recombination with unprecedented accuracy. However, even using the recently available ACBAR 2008 data and WMAP 3 year data, we can
get very informative restrictions on the parameters of the distortion models. In this Letter, we use the latest high resolution ACBAR data in combination with the WMAP data. The importance of the high resolution data seems to be quite obvious, since the major distortion on the CMB anisotropy power spectrum is related to the spatial scales on the order of $\Delta\sim 10$ Mpc, which is $\sim 3-5\%$ of the width of the last scattering surface. In addition to the models of delayed recombination \citep{Constraint_Recombination}, in this Letter, we consider the models of accelerated recombination, trying to constrain the cosmological model, based on the WMAP and the ACBAR data. We show that the WMAP data or the combination of the WMAP and the ACBAR data reveal some tendency in favor of  the ``accelerated'' recombination, while the constraint by the ACBAR data alone is more or less neutral.

\section{The deviation from the standard recombination process}
The production rate of extra resonance photons $n_{\alpha}$ is assumed to be \citep{Delayed_recombination,Ionization_history}:
\[\frac{d\,n_{\alpha}}{d\,t}=\epsilon_{\alpha}(z)\,H(z)\,n,\]
 where $n$ is the number density of atoms, $H(z)$ is the Hubble expansion rate at a redshift $z$, and $\epsilon_{\alpha}(z)$ is a parameter dependent on the production mechanism. Since the width of the recombination is very small in comparison to the horizon of the last scattering surface $L_{ls}$, the dependence of $\epsilon_{\alpha}(z) $ on $z$ can be parametrized as $\epsilon_{\alpha}(z_{rec}) +o(\Delta/L_{ls})$ . 
A simple parametrization using a constant effective value for $\epsilon_{\alpha}(z_{rec})\equiv\epsilon_{\alpha}$ with the WMAP data shows that  $\epsilon_{\alpha}<0.32$ \citep{Constraint_Recombination}.
Here $\epsilon_{\alpha}$ of negative values and positive values corresponds to accelerated recombination and delayed recombination respectively. The delayed recombination may be caused by the existence of the source for extra resonance photons \citep{Delayed_recombination,Ionization_history,Constraint_Recombination} and accelerated recombination may occur when there is clustering of baryonic matter on small scales \citep{Baryonic_recombination}. 
The possibility of accelerated recombination has not been given proper consideration, while the possibility of delayed acceleration has been investigated by many authors including \citep{Delayed_recombination,Ionization_history,Constraint_Recombination}. 

\begin{figure}[htb!]
\includegraphics[scale=.55]{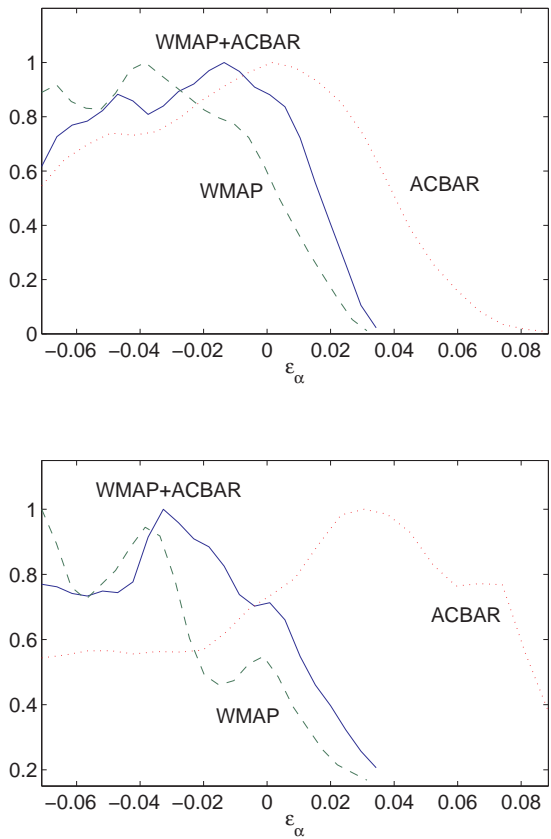}
\caption{the marginalized distribution (top), the mean likelihood (bottom). The normalization is chosen such that the peak value is equal to unity.}
\label{epsa}
\end{figure}

\begin{figure}[htb!]
\includegraphics[scale=.58]{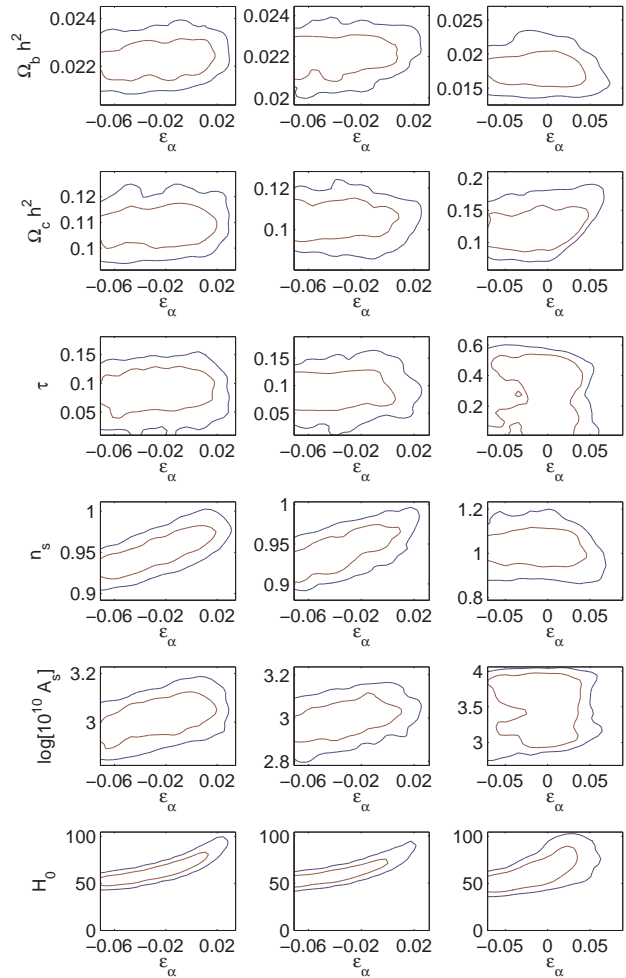}
\caption{The marginalized distribution on $\epsilon_{\alpha경부고속철도}$ vs 6 basic parameters, using the WMAP + ACBAR (left), the WMAP (middle), and the ACBAR (right).}
\label{epsa_2Dm} 
\end{figure}

\begin{figure}[htb!]
\includegraphics[scale=.58]{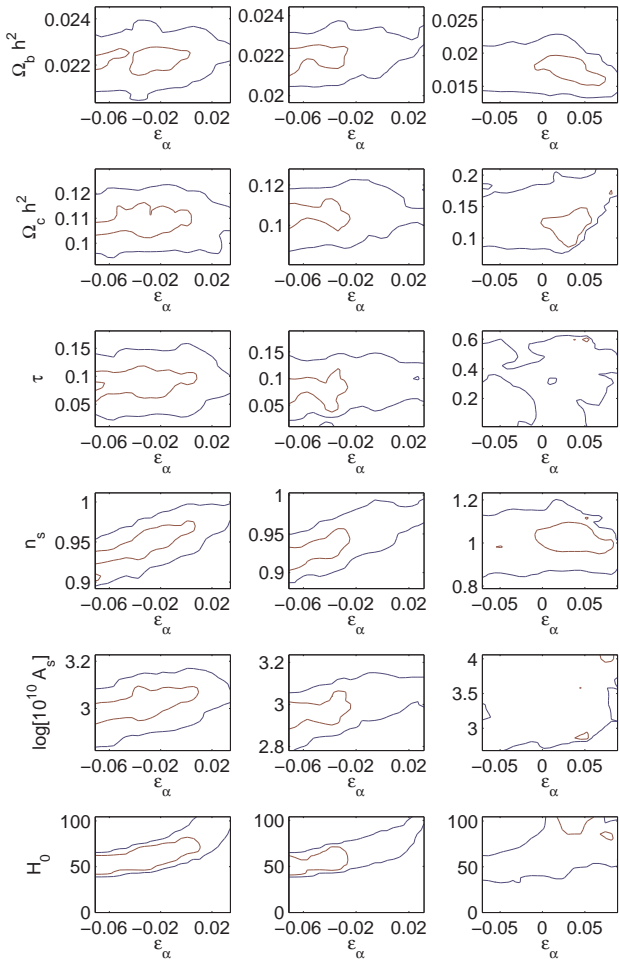}
\caption{The mean likelihoods of $\epsilon_{\alpha}$ vs 6 basic parameters, using the WMAP + ACBAR (left), the WMAP (middle), and the ACBAR (right).}
\label{epsa_2Dl}
\end{figure}

In this Letter, we have investigated the possibility of accelerated recombination as well, using the recent ACBAR 2008 data \citep {ACBAR2008} and WMAP 3 year data \citep{WMAP:3yr_TT}. 
Through small modifications to the widely used \texttt{RECFAST} code and the \texttt{CosmoMC} package \citep{CosmoMC}, we have included the parameter $\epsilon_{\alpha}$ in the cosmological parameter estimation, where we have explored the seven parameter space (six cosmological parameters +  $\epsilon_{\alpha}$).
We have assumed a constant effective value for $\epsilon_{\alpha}$ with the uniform prior $-0.07\le \epsilon_{\alpha}\le 0.3$, based on \citep{Ionization_history,Constraint_Recombination}. It should be noted that we have included $\epsilon_{\alpha}$ of negative values in the range $-0.07\le \epsilon_{\alpha}<0$, while the investigation by \citep{Constraint_Recombination} did not include $\epsilon_{\alpha}$ of any negative values for the prior.
The parameter space exploration was made through fitting the CMB power spectra of a flat $\Lambda$CDM model to the TT power of the ACBAR, and the TT, TE and EE power of the WMAP data.
The marginalized distribution shows the probability in the reduced dimension of parameter space, and the mean likelihood shows how good a fit can be expected \citep{CosmoMC}.  In the Fig. \ref{epsa}, we shows the marginalized distribution and mean likelihoods of $\epsilon_{\alpha}$, given the ACBAR and the WMAP data.
As shown in Fig. \ref{epsa}, $\epsilon_{\alpha}$ of negative values is as likely as that of positive values, implying the possibility of accelerated recombination.
When constrained by the WMAP data alone or by both of the WMAP and the ACBAR, $\epsilon_{\alpha}$ of negative values is more likely than that of positive values, although this is not in the c경부고속철도onstraint by the ACBAR data alone. 
This discrepancy in the constraint by the ACBAR alone may be attributed to the propagation of the uncertainty of other degenerate cosmological parameters to $\epsilon_{\alpha}$. The degeneracy between other cosmological parameters and $\epsilon_{\alpha}$ may be noticed in Fig. \ref{epsa_2Dm} and \ref{epsa_2Dl}, where we have plotted 1-$\sigma$ and 2-$\sigma$ contours of the marginalized distribution and mean likelihoods in the plane of $\epsilon_{\alpha}$ vs $\Omega_{b}h^2$, $\Omega_{c}h^2$, $\tau$, $n_s$, $\log[10^{10}A_s]$, $H_0$. We find that there exists some degeneracy between $\epsilon_{\alpha}$ and 
\{$n_s$, $\log[10^{10}A_s]$, $H_0$\}. From the Fig. \ref{epsa_2Dm} and \ref{epsa_2Dl}, we find again that $\epsilon_{\alpha}$ of negative values is more likely than that of positive values, when constrained by the WMAP alone or by both of the WMAP and the ACBAR.

The distortion on the CMB black body spectrum by the nonstandard recombination process ($|\epsilon_{\alpha}|<1$) is practically negligible in comparison with the distortion by the re-ionization history (see \cite{Antimatter} for details), and is consistent with the COBE FIRAS data constraint \citep{Fixen:FIRAS}.

\section{Conclusion and Summary}
We have investigated the distortion on the standard recombination process, using the ACBAR 2008 data and the WMAP 3 year data. 
We find that the constraint by the ACBAR and the WMAP data favors the possibility of accelerated recombination. 
As seen in Fig. \ref{epsa_2Dm} and \ref{epsa_2Dl}, there exists some level of degeneracy between $\epsilon_{\alpha}$ and other cosmological parameters. Partially due to these degeneracies, the constraint by the WMAP and the ACBAR are not powerfull enough to rule out the possibility of delayed recombination or the standard recombination at 1-$\sigma$ confidence level.
The polarization anisotropy is quite sensitive to the distortion \citep{Ionization_history}. Hence we will be able to impose tighter constraints on the distortion models, when the temperature and polarization data from the upcoming Planck satellite become available. 
\\\\

We are grateful to Antony Lewis for his guidance on the modification of the \texttt{CosmoMC} package.
We also thank the anonymous referee for helpful suggestion, which leads to the improvement of this paper. 
We acknowledge the use of the Legacy Archive for Microwave Background Data Analysis (LAMBDA) and ACBAR 2008 data.
Our numerical analysis was performed on the supercomputing facility of the Danish Center for Scientific Computing.
This work was supported by FNU grants 272-06-0417, 272-07-0528 and 21-04-0355.

\end{document}